\documentstyle[12pt,epsf]{article}
\begin{document}
\newcommand{\be}{\begin{equation}}
\newcommand{\ee}{\end{equation}}
\newcommand{\ba}{\begin{eqnarray}}
\newcommand{\ea}{\end{eqnarray}}
\newcommand{\pa}{\partial}
\newcommand{\f}{\frac}
\newcommand{\st}{\stackrel}
\newcommand{\tk}{\tilde\kappa}
\newcommand{\ep}{\epsilon}

\begin{center}
{\huge\bf Oscillating structure of $\gamma$-bursts and their possible origin}

\vspace*{0.5cm}
{\large\bf S.S. Gershtein,}\\
{\bf Institute for High Energy Physics,}\\
{\bf Protvino, Moscow region,}\\
{\bf 142284 Russia.}\\
{\bf E-mail: gershtein@mx.ihep.su}
\end{center}

{\small 

As it is well-known that the  hydrodinamic collapse of the
massive star iron core should lead to the production of a hot neutron star.
The assumption is made that the thermonuclear burning of the
envelope matter, accreting onto the hot neutron star,
can proceed in the oscillatoric regime (analogously to that happens 
during heat explosion
of the carbon-oxigene cores of stars with smaller masses).

Local density oscillations in the vicinity of the neutron star surface can generate
shock waves, in which the stratification of the electron-positron
plasma from the rest of the matter 
can happen due to the light preasure. In the case of the spherically symmetric
collapse  of the compact star it can lead to  the production of 
the expanding relativistic
fireball shells with characteristic oscillation time of  $\sim 10^{-2}$~s, 
observed in the cosmological $\gamma$-bursts (GRB), can occur. 

It is pointed out that nonrotating massive Wolf-Rayet's (WR) stars could be the source 
for the GRB, whose collapses, according to a number of observations, can happen
without any noticeable ejection of the envelope.}

\vspace{1cm}
\section{Introduction. Fireball model}

\noindent
Observed $\gamma$-bursts (GRB --- Gamma Ray Burst) is extremely interesting
and still unexplained phenomenon (see reviews [1-4] and refs. therein).

Optical identification of the $\gamma$-bursts with "host" galaxies has proved
that, at least, a part of them occurs in the galaxies with red shift
of $Z\geq 1$, i.e. has the cosmological origin. This agrees well with
a fully isotropic distribution of GRB over the sky and with 
statistic distribution of the burst events over their intensity. 

Optical identification of GRB has allowed one to determine the distance to them 
and establish that  a huge energy of $\sim 10^{52}\div 10^{54}$~erg.
in the $\gamma$-range (30-500~keV) is emitted in this phenomenon. Many of 
observed GRB characteristics are explained in the framework of the  
``fireball'' model [5-7], i.e. in terms of the electron-positron cloud, expanding
with ultrarelativistic velocities. Ultrarelativistic velocities of the expansion
(naturally appearing in the electron-positron plasma)~[6],  allow one to solve
the problem of the GRB source compactness [6-8] and match the nonthermal
 GRB spectrum  with short characteristic time of the GRB variability 
 $(\delta t\sim 10$ms.)\footnote{It is very important here that the bound on 
the number of baryons contained, which has to be small enough [9,10],
is a condition of the ultrarelativistic expansion of lepton-photon plasma}.
Basing on the fireball model one has succeeded to explain both the observed effects of 
the long-lasting optical GRB afterglow, which appears as a result 
of the interaction between relativistic expanding fireball and inter-stars medium
[11], and effect of the early afterglow, which intersects in time with   
GRB of high duration [12-14]. Thus, the model of the relativistic
fireball containing a small number of baryons allows to agree the observed
GRB characteristics and explain accompanying phenomena. However, some of questions
remain unresolved yet:

1. The mechanism of the fireball production.

2. Large energy in the fireball.

3. The presence in some GRB's of large number $(\sim 10^2\div 10^3)$ 
of pulsation of the $\gamma$ emittion intensity with
characteristic time of $\delta
t\approx$10~ms. We believe that this effect could serve as
basic key to resolve the $\gamma$-bursts puzzle. In this paper
we appeal the attention to the fact that the oscillations 
of the  $\gamma$-quantum flux can naturally appear during the
hydrodynamic collapse of some compact,
massive, and nonrotating stars at final stage of their evolution. 

\section{The possibility of the oscillator burning
of thermonuclear fuel during the process of hydrodinamic
collapse} 

It is well-known that the production of sufficiently large iron core 
in the process of the star evolution is a reason of the hydrodinamic collapse
of massive stars with masses  $M\geq 10M_\odot$. 
In this case the star core, having exhausted the source of thermonuclear energy, 
tends to compress and heat. The resulting increase of preasure, however,
is unable to stop the compression, since the thermal energy is spent to
the endothermal reaction of  the iron nucleus decay and further - to the core
neutronization. As a result, the core compression is transferred 
to the catastrophic hydrodinamic collapse, which is followed by
the production of the hot neutron star. According to the idea by Fowler and Hoyle [16], 
the accretion of the nuclear fuel, which is left in the star envelope, onto
the hot neutron star leads to its explosion and pollution manifesting itself
as the burst of  supernova. However, selfconsistent hydrodinamic calculations 
did not prove this assumption. It turned out that consequent account for
the neutrino emittion leads to the delay of the collapse, which stops only when
the core matter becomes nontransparent to the neutrino radiation.
As a result of this delay, the burning of the accreting nuclear fuel
occurs in a deep gravitational potential and, thus, the emitted thermonuclear
energy is insufficient for the envelope ejection [17] (see, for instance, review
[18] and refs. therein). The shock wave, occurring during the collapse delay,
dumps only a small fraction of the envelope with the energy about $10^{49}$erg [19],
that is two orders of magnitude less than a characteristic energy of the
supernova explosion, $10^{51}$erg. Thus, this phenomenon was named as 
 "soundless" or silent collapse. Further attempts to explain the supernova bursts
at the spherically symmetric collapse of massive stars had not led to a desirable
result [20]  and now the most of specialists tend to the idea that the observed
supernovas bursts with massive early-supernovas are anyway connected with 
effects of the collapsing star rotation: magnetic preasure onto the envelope,
the Relay-Taylor instability, or breakage of the neutron star into two components 
(see, for instance, [21-22]).

However, one should  take into account that the burning of the thermonuclear 
fuel during its accretion onto a hot neutron star can have the oscillating
behavior. This effect is well-known and manifest itself in the consideration
of the final evolution stage of stars with small masses, $3M_\odot \leq M\leq
10M_\odot$. In such stars the oxigene-carbon core with degenerated electronic gas
is produced as a result of the evolution. 
In this case the thermal explosion in the degenerated star core is the reason of 
the instability, when the core mass achieves the value close to the Chandler limit
[23]. In this case the oscillating character of thermonuclear fuel burning could be
easily understood, if one takes into account a relatively small calorific power
of this fuel with carbon-oxigene etc. content.
The energy emitted during the thermal explosion leads to elimination of the
degeneracy and increase of the  thermal preasure resulting in the
star expansion. As a result	of the expansion the star temperature is decreased.
This leads to the consequent compression of the star and enhancement of the
thermonuclear burning which, in its turn, leads to the expansion, etc.
All said above can be illustrated in Fig. 1, where one can see the results
of calculations [24]. Observed oscillation not only retained, but  
have been enhanced due to the account for convection, thus leading to
delayed detonation with explosion energy $\sim 10^{51}$erg 
(report by V.S.Imshennik, seminar devoted to the memory of
 S.I.Syrovatsky, March 2, 2000). 

It is possible that such oscillations appears
as well in the layers of the thermonuclear fuel accreting onto a hot neutron star.
In contrast to the case discussed above, they can have only local character,
evolving in layers, adjacent to the surface of the hot neutron star. 
The period of these oscillations can be estimated by the arguments of
dimension: 
\be
\tau\sim \frac{1}{\sqrt{G_N\bar\rho}},
\ee
where $G_N$ is the gravitational constant,  $\bar\rho$ is the matter density
in the vicinity of the neutron star. According to the calculations 
(see Fig. 2), $\bar\rho \sim
10^{11}$g/cm$^3$. Thus, the period of oscillations turns out to be equal to
\be
\tau\sim 10^{-2}s,
\ee
that intriguesly coincides with oscillation period of  
the $\gamma$-quantum flux in some GRB\footnote{It should be noted that
the conditions for oscillation excitation [25] are also realized
in the hot neutron stars, produced as a result of the iron core collapse. 
However, their frequency is, at  least, two orders of magnitude
more.} Density and temperature oscillations close to the surface of
hot neutron star have to generate diverging shock waves in the
surrounding envelope (with decreasing density vs. radius increase), repeating
with the oscillation frequency.

\section{The possibility of the electro-positron plasma stratification}

\noindent
One of the most important effects, which should be taken into account
in description of shock waves passing through the star envelope,
is a possible stratification of the electron-positron plasma, which 
happens without violation of the electroneutrality of ordinary matter
(nuclei and electrons, which compensate their electric charge). Such
stratification is possible under the condition of light preasure
in propagating shock wave, as the Eddington limit for electron-positron 
plasma is 3600 times lower than that for an ordinary matter with nuclei $A\simeq 2Z$. 
So, the electron-positron plasma, when exiting to the star surface, 
will contain a relatively low concentration of baryons (that is just needed
to agree the fireball model and observed data).
It should also be noted that under the condition of the rarefacted star atmosphere 
the equilibrium electro-positron plasma can appear at relatively low temperatures, since
under these conditions at $kT<<m_ec^2$:
\ba
n_{e^+}\approx n_{e^-}&\approx& \f{1}{(2\pi^3)^{1/2}}
\left (
\f{m_ec}{\hbar}\right )^3e^{-1/x}x^{3/2}\nonumber \\ [-0.2cm]
\\ [-0.2cm]
x&=&\frac{kT}{mc^2}<<1 \nonumber
\ea
while in the dense matter the positron concentration
will be proportional to $\exp-\left (\frac{2mc^2}{kT}\right )$ (see, for instance, 
[26]). The fact that sufficiently high temperatures $(kT\geq m_ec^2)$ are achieved
close to the center of the collapsing star is proved by the  presence of the process
of explosive nucleosynthesis of the  $^{56}Ni$ nuclei with subsequent
production of $^{56}Co$, as it comes from the observation data on SN$1987$~[21,27]. 
(According to the calculations~[19], in the region of neutrinosphere
 $kT\simeq 5,6$~MeV). 

When leaving the star atmosphere the expanded cloud of electron-positron plasma
inevitably gains the ultrarelativistic character (see [6]).
This, as it is well-known, leads to variations of the
$\gamma$-radiation momentum, received by a remote observer. 
\be
\delta t\simeq \frac{R}{2c\Gamma^2},
\ee
where $R$ --- the cloud radius, $\Gamma=(1-v^2/c^2)^{-1/2}$ --- the Lorentz factor,
corresponding to expansion velocity, $v$. It is evident that
the oscillation of the  $\gamma$-quantum flux will be observed if
\be
\delta t\leq \tau.
\ee
In opposite case $(\delta t>> \tau)$, the oscillations in the $\gamma$-quantum
flux for a remote observer are smeared. This helps to explain the fact that 
there are no observed oscillations in some GRB. 
Stratification of the electron-positron plasma from
ordinary matter should lead to situation, when each oscillation 
in the vicinity of the neutron star will produce shock waves in the form of
two shells expanding with different velocities. Here, it could happen that
the electron-positron shell, emitted in the following oscillation
can overtake that one containing baryons and emitted in previous oscillation.
Thus, there can be an interaction of shock waves
inside the fireball itself~[12,13,28,29,30]. 

Ultrarelativistic character of the fireball expansion $(\Gamma\sim 10^2)$ allows to
conclude that at observed values of  $\delta t\sim
10^{-2}$s. the fireball size can be large enough (and the plasma density is low enough,
correspondingly), to consider the fireball as a  ``thin'' source. 
It allows to explain the nonthermal (power) spectrum of GRB.
``Intrinsic'' shock waves, producing the fireball, can also generate 
high energy particles by means of well-known mechanism of acceleration.
This can explains the observation in some GRB the high energy $\gamma$-quanta 
(up to 18 GeV).

\section{Possible progenitors of GRB}

\noindent
The scenario of the oscillation origin in GRB developed above confines
the class of object, which could be the progenitors of GRB. 

First, these should be sufficiently massive stars with masses $M > (15-20)M_\odot$. 
Quite stage of such star evolution has to be ended within a time period
about $10^6$ years or less. Star mass, large enough, is also required to explain the 
GRB energy.

Second, these should be nonrotating  (or with low angular velocity) stars. It seems
that stars with high angular velocity should explode due to the effects connected with
their rotation, as ordinary supernovas of the II-type with ejection of relatively
massive envelope~[31]. 

And third, these should be compact stars devoided of extent hydrogen ad, probably,
in part helium envelope, which is able to prevent the outer ejection of the 
electron-positron plasma due to the processes of positron annihilation. 

The stars of the Wolf-Rayet type (WR) meet all these requirements --- they are 
the most massive compact start, which have lost	almost all their hydrogen and,
in part, helium envelope during their evolution. It is possible that, namely,
due to the loss of the balk of their envelope these stars have lost 
their rotatory impulse. Anyway, the rotation is observed only for
15\% of the  WR-stars [32]. In the studies by
A.M.~Cherepashchuk et al. (see [33] and refs. therein) it was found that
one can neglect the decrease of the WR-star masses in the process of their
further evolution (which is caused by the stellar wind). It allows one to compare
the masses  of the WR-stars and their 'Ž-nuclei with masses of the relativistic
objects (neutron stars and ``black holes''), for which the WR-stars  are
the progenitors. 
Basing on the masses measurement of the X-ray sources in double systems 
A.M.~Cherepashchuk has drawn the extremely important conclusion that
the distribution of the X-ray sources masses has clear bimodal character. 
There is a mass gap between the neutron stars --- pulsars, whose masses
are ranged in the narrow band $(1-2)M_\odot$ with average mass  $(1.35\pm 0.15)M_\odot$,
and masses of candidates to black holes, which are distributed in the range
of $(5\div 15)M_\odot$ and have average mass of
$(8\div 10)M_\odot$. The bimodal mass distribution and
the presence of the gap serves as the indication to different origin of these objects.
As for the massive candidates to black holes, the correlation (discovered by
A.M.~Cherepashchuk [33]) between their masses
and masses of the WR stars, which are in the range 	of $(5- 55)M_\odot$ and for which 
the average value of their CO-nuclei is  $(8\div 12)M_\odot$, close to the average value
of masses of the observed candidates to black holes, seems to be very important
to understand their origin.
Thus,
there are arguments to assume that, at least, some of the WR stars collapse
into massive objects through the ``soundless'' collapse without 
any significant ejection of their envelope. In the first turn it concerns
more evolved WC stars with reach content of C-nuclei in their envelope 
(produced due to the thermonuclear burning of helium) and with average mass of
$13,4\;M_\odot$. 
The data presented in [33] are the strong
argument in the favour of the assumption formulated by P.Conti in 1982~y. [34] that
the WR-stars more often disappears in the form of the ``whimper'', rather than
explosion \footnote{It is possible that some WR-stars are early supernovas~1b. 
(I would like to acknowledge this remark by V.S.~Imshennik).} 

The compact structure of the WR-stars allows one to assume
that the electron-positron plasma shells, which appears due to the
stratification  in shock waves, can leave the star surface and even a small
part of the large gravitational energy emitted in the massive star collapse
can explain the GRB energy. 

\section{The GRB energy}

To determine the GRB energy it is necessary to have
self-consistent hydrodinamic calculations of the process of soundless
massive star collapse with the  account for the possibility of the
$e^+e^-$-plasma stratification from the rest of the matter. However,
one can try to estimate a possible GRB energy from physical
(though, not very reliable) arguments. 
If as the result of the collapse into the hot neutron star with mass $M=1,5M_\odot$
the gravitational energy $\epsilon\approx
5\cdot 10^{53}$erg. is emitted in the form of the neutrino radiation, then for masses
$M=(15\div 20)M_\odot$ (in the case without the envelope ejection) one can
expect the energy emittion about $\epsilon\simeq 10^{56}$erg.\footnote{It should
noted that a hot neutron star should be stable up to the mass of $M_{NS}
\approx 70 M_\odot$~[35].}.
So, to provide the GRB energy $\sim 10^{53}$erg. it is sufficient
 to have $(e^+e^-)$-plasma ejection accumulated  
$\sim$0,1\% of emitting gravitational energy. Analogous estimate
one can obtain using (taking some risk) the calculation results of the hydrodinamic
collapse of the iron-oxygene star core~[18]. Though the shock wave
generated in this process is subjected to the attenuation due to neutrino
radiation and its power is not sufficient to explain the supernova explosion, 
nevertheless,
the energy of the emitted shell can be  $\sim 10^{49}$erg.  For the
obtained velocity of the shell expansion of $v\sim 1,5\cdot 10^3$km/s the mass
of the ejected shell is $\Delta M\simeq 0,44 M_\odot$. Were such mass
being ejected in the form of the  $e^+e^-$-plasma, the energy of $8\cdot 10^{53}$erg.
should be emitted during its subsequent annihilation. The comparison given here
is not proved well, but it gives an idea on the possible effect value. 
Thus, the bulk of the GRB energy in the mechanism considered  has 
the gravitational origin. The heating of the collapsing star leads to
the production of the dense and hot $e^+e^-$-plasma, and energy emitted 
in oscillatoric burning of the thermonuclear fuel is spent to generation of
shock waves pushing the $e^+e^-$-plasma beyond the star. 

One should also take into account the possible additive energy to the
expanding $e^+e^-$-plasma due to neutrinos and antineutrinos radiated
during the collapse process (since, during their scattering  on the
electrons and positrons they can pass to laters the bulk of their energy). 
The question  of $\gamma$-radiation spectra from GRB requires special treatment.
It is quite possible that the absence of the  511keV line from the 
$e^+e_-$-annihilation at the rest is connected with the ultrarelativistic
fireball expansion. 

\section*{Discussions}

\noindent
On the basis of the observation data [33] there are forcible arguments
to assume that massive, compact, and nonrotating stars of the Wolf-Rayet type
[32] are subjected to the relativistic collapse without any significant
ejection of their envelope.	This assumption agrees well	with the fact that
within the hydrodinamic calculations one fails to obtain the envelope ejection 
sufficient to explain the supernova bursts.	The gravitational energy emitted
during the relativistic collapse of such objects can be about 
$10^{55}\div 10^{56}$erg. 

The hypothesis developed in this paper is that the burning 
of the thermonuclear fuel accreting onto a hot neutron star 
can proceed in the oscillatoric regime, which generate the shock waves,
which, in their turn, push out the $e^+e^-$-plasma 
outside the star surface (in the presence of its stratification).
This hypothesis qualitatively explains
the origin of the relativistic fireball with a low baryon content and oscillations
observed in GRB (moreover, the oscillation period is explained quantitatively
in the order of magnitude). The duration of  GRB
($\sim 20$s) agrees with the time of the outer envelope accretion onto the neutron
star and time of its cooling. A number of observed data, provided by B.~Paczynski,
in particular the indication that GRB happen in the regions of intensive
star production [36], tell in the favor of the fact that  
the WR-stars can be the progenitors of GRB.  

According to the hypothesis suggested in this paper, the collapse of the 
WR-star and ejection of the $e^+e^-$-plasma happen spherical symmetrically.
Successful description of the optical afterglow at time-period $\sim 200$~days
[37], obtained in the framework of this hypothesis, proves in favour
of the spherical symmetry of a number of the $\gamma$-bursts. 
(However, there are some indications to the fact in that jets can appear in some 
of GRB.) 

In the conclusion the author would like to thank
G.V.~Domogazky, A.M.~Dyhne, A.A.~Logunov, V.S.~Imshennik, D.K.~Nadezhin, 
K.A.~Postnov for the stimulating interest to this work and valuable remarks. Special thanks
to A.M.~Cherepashchuk for letting me know his work and data on the WR stars. 

This work is supported, in part, by the RFBR grants
99-02-16558 and 00-15-96645.

\newpage
\setlength{\unitlength}{1mm}
\begin{figure}[ph]
\begin{center}
\begin{picture}(100, 80)
\put(0, 0){\epsfxsize=11cm \epsfbox{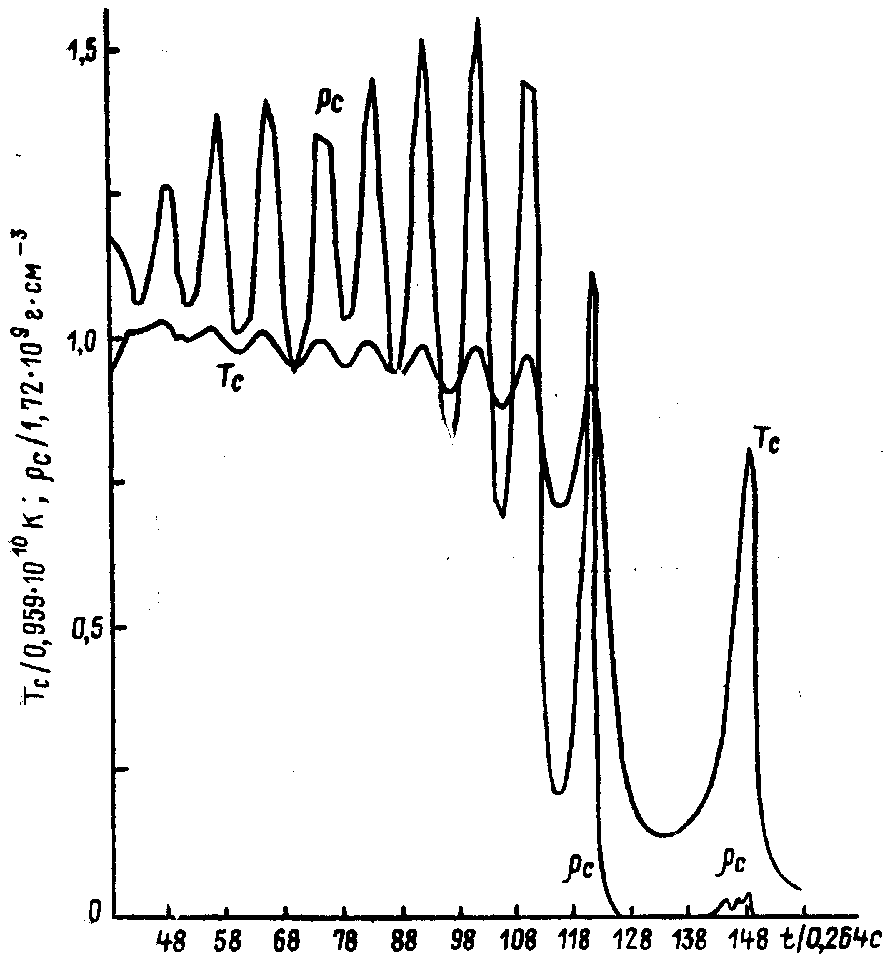}}
\end{picture}
\end{center}
\caption{Time dependence of the central density and temperature in the process
of the carbon burst. The pulsing regime of the carbon burning with subsequent 
separation of star core fragments is clearly seen~[24,21].}
\label{fig1}
\end{figure}

\newpage
\setlength{\unitlength}{1mm}
\begin{figure}[ph]
\begin{center}
\begin{picture}(100, 80)
\put(0, 0){\epsfxsize=11cm \epsfbox{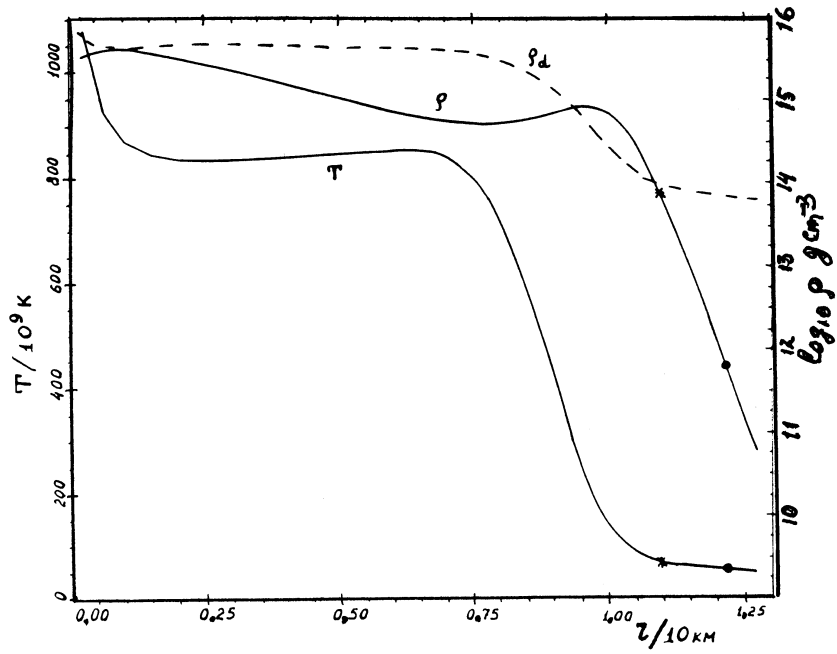}}
\end{picture}
\end{center}
\caption{The density and temperature distributions over the hot neutron star
surface (dashed line corresponds to the degeneracy density). The location of the 
neutrino photosphere is shown by the star marker, the black marker shows the 
boundary of the neutron core~[19].}
\label{fig2}
\end{figure}

\end{document}